# Ultra-wideband MRE of the human liver and spleen for viscoelastic model identification in hepatic inflammation

Jakob Schattenfroh[1], Michael Fedders[1], Steffen P. Häseli[1], Tobias Puengel[2], Yanglei Wu[1], Tom Meyer[1], Hossein S. Aghamiry[1], Tim Ricken[3], Münevver Demir[2], Frank Tacke[2], Jing Guo[1], Rolf Reiter[1,4], Ingolf Sack[1]

1. Department of Radiology, Charité – Universitätsmedizin Berlin, Germany
2. Department of Hepatology and Gastroenterology, Charité – Universitätsmedizin Berlin, Campus Virchow-Klinikum (CVK) and Campus Charité Mitte (CCM), Germany
3. Institute of Structural Mechanics and Dynamics in Aerospace Engineering, Faculty of Aerospace Engineering and Geodesy, University of Stuttgart, Germany
4. Berlin Institute of Health at Charité – Universitätsmedizin Berlin, BIH Biomedical Innovation Academy, BIH Charité Digital Clinician Scientist Program, Berlin, Germany

**Corresponding Author**

Ingolf Sack, Department of Radiology, Charité – Universitätsmedizin Berlin, Chariteplatz 1, 10117 Berlin, Germany, ingolf.sack@charite.de

## Abstract

Magnetic resonance elastography (MRE) is an established technique for the noninvasive assessment of liver fibrosis. However, conventional abdominal MRE typically operates within a narrow frequency range of 40-60 Hz. Lower frequencies remain largely unexplored, particularly with regard to hepatic inflammation. We developed in vivo ultra-wideband MRE covering 5-80 Hz to investigate frequency-resolved viscoelastic dispersion of the liver and spleen and to identify biomechanical markers sensitive to hepatic inflammation.

Following phantom validation, nine healthy volunteers and nine patients with clinically diagnosed inflammatory liver disease were prospectively examined at 12 excitation frequencies, three motion-encoding components, and 32 vibration phases. Spatiotemporal phase unwrapping and frequency-adaptive wavefield preprocessing enabled reconstruction of shear wave speed (SWS), penetration rate (PR), and loss angle (ϕ). Six rheological models with two to three degrees of freedom were evaluated to describe liver and spleen dispersion. MRI-based biomechanical studies of the liver have employed a broad range of constitutive models, and the resulting material parameters depend substantially on model choice and acquisition protocol.

Model-independent parameters showed the strongest inflammation-associated changes at frequencies below 20 Hz: ϕ increased by 63% ($p<0.001$), PR decreased by 37% ($p=0.003$), and SWS increased by 8% ($p=0.008$), indicating predominantly dissipative rather than stiffness-related mechanical changes and an inflammation-associated shift toward fluid-like behavior with minor stiffness changes in the lower frequency regime. In the full frequency range of 5-80 Hz, the rheological springpot model with serial dashpot provided the best fit and revealed distinct dispersion functions for liver and spleen. In patients, springpot elastic modulus increased (101%, $p=0.001$), while viscosity and springpot power-law exponent decreased (52%, $p=0.002$ and 58%, $p<0.001$) suggesting a shift from soft-fluid to stiff-solid liver properties.

Ultra-wideband MRE revealed that clinically diagnosed inflammatory liver disease is associated with property shifts toward stronger dissipation and fluid-like behavior at low frequencies while displaying solid-like behavior at higher frequencies. Ultra-low-frequency MRE, particularly at frequencies below 20 Hz, may provide a diagnostic window into inflammation-associated liver viscoelasticity without full rheological modeling.

## Introduction

The viscoelastic properties of organs and soft tissues are increasingly recognized as sensitive biomarkers of disease. Magnetic resonance elastography (MRE) [1] noninvasively maps these properties by imaging the propagation of externally induced shear waves, thereby providing spatially resolved measures of tissue biomechanics across abdominal organs [2]. MRE has become a clinical tool for the noninvasive assessment of liver fibrosis, e.g. in metabolic dysfunction-associated steatotic liver disease (MASLD) [3]. Spleen stiffness has emerged as a key surrogate marker of portal hypertension [4]. The viscoelastic behavior of the liver and spleen also provides markers of inflammation [5–7] and of portal pressure- and blood-flow-related disease processes [8–12].

Viscoelasticity integrates viscous, i.e. dissipation-related, and elastic, i.e. stiffness-related, material responses to external deformation, both of which can vary substantially with the frequency at which tissue is mechanically excited [13,14]. For example, in vivo biological soft tissues such as the brain can exhibit very low stiffness values of 5-25 Pa at 1 Hz [15], low to intermediate values of approximately 250 Pa and 840 Pa at 5 Hz and 10 Hz, respectively [16], and substantially higher values exceeding 1 kPa above 10 Hz [17]. The corresponding shear wavelengths in the order of only 7-10 centimeters at frequencies below 10 Hz are much shorter than would be expected for elastic, non-dispersive materials. This pronounced frequency dependence of stiffness (a factor of 20 within a decade of Hz) has motivated terms such as "superviscous brain" [17] or "liquid liver" [18].

Despite recent progress in the in vivo application of wideband MRE in the liver [14,19], there is still no consistent measurement framework for viscoelastic dispersion of in vivo abdominal organs that covers a continuous frequency range from ultra-low frequencies (1-5 Hz) to the clinical range (approximately 60 Hz or higher). Consequently, it remains unknown whether specific frequency windows provide enhanced sensitivity to different pathological processes, including steatosis, inflammation, fibrosis, or elevated portal pressure [20]. The detection of hepatic inflammation remains a challenge for noninvasive medical imaging. Liver biopsy, despite its invasiveness, sampling variability, and procedural risk, therefore, often remains necessary for a definitive diagnosis and grading of hepatic inflammation [21]. Although iron-corrected T1 (cT1) mapping provides an MRI-based proxy for hepatic disease activity in selected liver disease conditions [22], it is not specific to inflammation and may therefore be complemented by MRE protocols tailored to the biomechanical cascade of inflammation [23,24].

Hallmarks of inflammation associated with biomechanical changes are vascular leakiness, interstitial fluid accumulation, immune-cell infiltration, and matrix secretion of polar proteoglycans and glycosaminoglycans following acute or chronic injury [25–27]. In contrast to fibrosis, which predominantly affects tissue stiffness by accumulation and crosslinking of extracellular matrix proteins [20], inflammatory processes are expected to influence solid-fluid interactions and increase the dissipative behavior of soft tissue. Poroelastic models of multiphasic soft tissues predict that solid-fluid interactions become increasingly important at low excitation frequencies, where the fluid phase can move independently from the solid matrix [28–31]. Liver tissue has previously been represented as a biphasic porous medium in which sinusoidal perfusion, pressure gradients, and tissue remodeling are mechanically coupled [32]. Consistent with this concept, the sensitivity of MRE-derived stiffness to fibrosis is greatest at higher excitation frequencies [33,34] in both the liver and spleen [35]. The loss angle of the shear modulus [36] and the related viscoelastic power-law exponent [37] have been identified as promising biomarkers of hepatic inflammation, consistent with the concept that loss angle reflects tissue fluidity [23]. Together, these findings suggest that frequency-resolved measurements of viscoelastic dispersion, particularly at very low frequencies, could improve the detection and characterization of hepatic inflammation [38].

We developed ultra-wideband MRE for the human liver and spleen covering a continuous frequency range from 5 to 80 Hz. We assessed the frequency-specific sensitivity of MRE parameters, particularly loss angle, to hepatic inflammation in vivo from ultra-low (5 Hz) to high vibration frequencies (80 Hz). The continuous frequency coverage over four octaves allows both grouped frequency analysis according to the principles of high-resolution tomographic MRE [2], as well as rheological model fitting [14] to identify candidate viscoelastic parameters for detecting hepatic inflammation. Following technical validation in a phantom that mimics liver viscoelastic dispersion [18], novel spatiotemporal phase unwrapping [39] and frequency-adaptive wavefield preprocessing [16] were applied in healthy volunteers and patients with clinically diagnosed hepatic inflammation to evaluate the sensitivity of model-independent viscoelastic parameters and rheological model parameters to inflammatory liver disease.

## Methods

### Study Design and Participants

This cross-sectional observational study included two groups: (i) healthy volunteers without known hepatic or abdominal pathology (N=9; 2 female; mean age=28±5 years; BMI=23.4±2.2 kg/m$^2$) and (ii) patients with clinically confirmed inflammatory liver disease (N=9; 4 female; mean age=54±15 years; BMI=27.2±6.6 kg/m$^2$). Patient diagnoses are summarized in Supplemental Table 1. All participants provided written informed consent. The study was approved by the institutional ethics review board.

### MR Elastography Acquisition

All MRE examinations were performed on two clinical 3T MRI systems (MAGNETOM Lumina and MAGNETOM Skyra, Siemens Healthineers, Erlangen, Germany) at the same institution and using identical MRE acquisition protocols and actuator hardware.

Continuous harmonic vibrations were generated by four pressurized-air drivers, with two drivers placed laterally over the rib cage and two drivers placed dorsally, all of which were secured with a Velcro belt [2]. The drivers were connected to a multi-channel pressure control unit synchronized with the MRE sequence. The continuous frequency range from 5 to 80 Hz was acquired in two acquisitions: a low-to-mid (5-35 Hz, 5 Hz increments) and a mid-to-high-frequency regime (40-80 Hz, 10 Hz increments).

MRE data was acquired using a single-shot spin-echo echo-planar imaging (SE-EPI) sequence with flow-compensated motion-encoding gradients (MEGs). Wave data were consecutively acquired across 32 or 8 dynamics of the vibration cycle (phase offsets), three Cartesian axes and twelve frequencies [40]. 32 phase offsets per frequency were acquired in the low-to-mid-frequency regime to stabilize phase unwrapping with additional temporal information while 8 phase offsets per frequency were acquired in the standard mid-to-high regime. Examinations were performed during free breathing with retrospective motion correction [41]. The imaging volume covered the liver and spleen simultaneously. Imaging parameters are further listed in Table 1.

**Table 1:** Wideband MRE acquisition parameters for the low-mid and mid-high frequency regimes. The values in the parentheses represent the mid-high frequency acquisition.

| Acquisition Parameter | Value |
|---|---|
| **Field-of-view ($mm^3$)** | 360 x 252 x 25 |
| **Matrix** | 120 x 84 x 5 |
| **Voxel size ($mm^3$)** | 3.0 x 3.0 x 5.0 |
| **TR / TE (ms)** | 805 / 51 |
| **Receiver Bandwidth (Hz/pixel)** | 1736 |
| **Acceleration** | GRAPPA, R=2 |
| **Driver Frequencies (Hz)** | 5, 10, 15, 20, 25, 30, 35 (40, 50, 60 ,70, 80) |
| **Phase Offsets** | 32 (8) |
| **MEG Amplitude (mT/m)** | 34 |
| **MEG Slew Rate (mT/m/ms)** | 125 |
| **Total MRE acquisition time (min)** | 9:17 (1:29) |

**Data Processing and Inversion**

The wideband abdominal MRE framework consisted of two technical developments to enable viscoelastic dispersion mapping of the liver and spleen from 5 to 80 Hz: (i) dedicated phase unwrapping that recovers reliable wavefields at low-frequency excitation, and (ii) dispersion-sensitive high-pass filtering to isolate shear waves from compression waves across a wide range of wavelengths adapted from the publicly available k-MDEV inversion (bioqic-apps.charite.de) [42,43]. Each development is described in further detail below. Liver and spleen ROIs were manually drawn on MRE magnitude images by an experienced reader. Because higher-frequency shear waves undergo stronger attenuation and the coverage of sufficient wave amplitude decreases with frequency, regions with insufficient wave propagation may be excluded separately at each frequency [2,44].

**(i)** **Spatiotemporal Phase Unwrapping:** Phase unwrapping was performed using novel Iterative Laplacian Phase Unwrapping (ILPU) [39] and is illustrated in Figure 1. First, a voxel-wise phase-quality map (Fig. 1B) was derived by quantifying the inconsistency between local phase differences in the original wrapped phase image (Fig. 1A) and those obtained after single-step global Laplacian phase unwrapping. The phase-quality map was thresholded and combined with the manually defined liver and spleen ROIs to generate a binary mask of regions containing reliable phase information (Fig. 1C). ILPU [38] was then applied separately within each connected organ region (Fig. 1D), thereby minimizing phase inconsistencies between adjacent anatomical structures and reducing phase-aliasing errors at organ boundaries. Finally, voxels within the abdominal ROI but outside the reliable-phase regions were recovered by Laplacian phase unwrapping with Dirichlet boundary conditions, using the regional ILPU solutions as fixed boundary values. This step produced the globally unwrapped phase image (Fig. 1E). Further details are provided in Supplemental Note S2.

**(ii) Frequency-adaptive bandpass and directional filtering:** After temporal Fourier transformation of the phase-offset series, the complex wavefield at the fundamental frequency was extracted [42]. A frequency-adaptive linear radial high-pass filter suppressed low-wavenumber compression wave contributions while preserving shear waves [42]. To accommodate the wavelength range across 5-80 Hz, the slope was defined as $s(f) = [1 + 0.4 \cdot (1 - f/20)^4]^{-1}$ for f<20 Hz and $s(f) = 1$ otherwise [14]. A cosine-squared directional filter then decomposed each wavefield into mono-directional planar shear wave components.

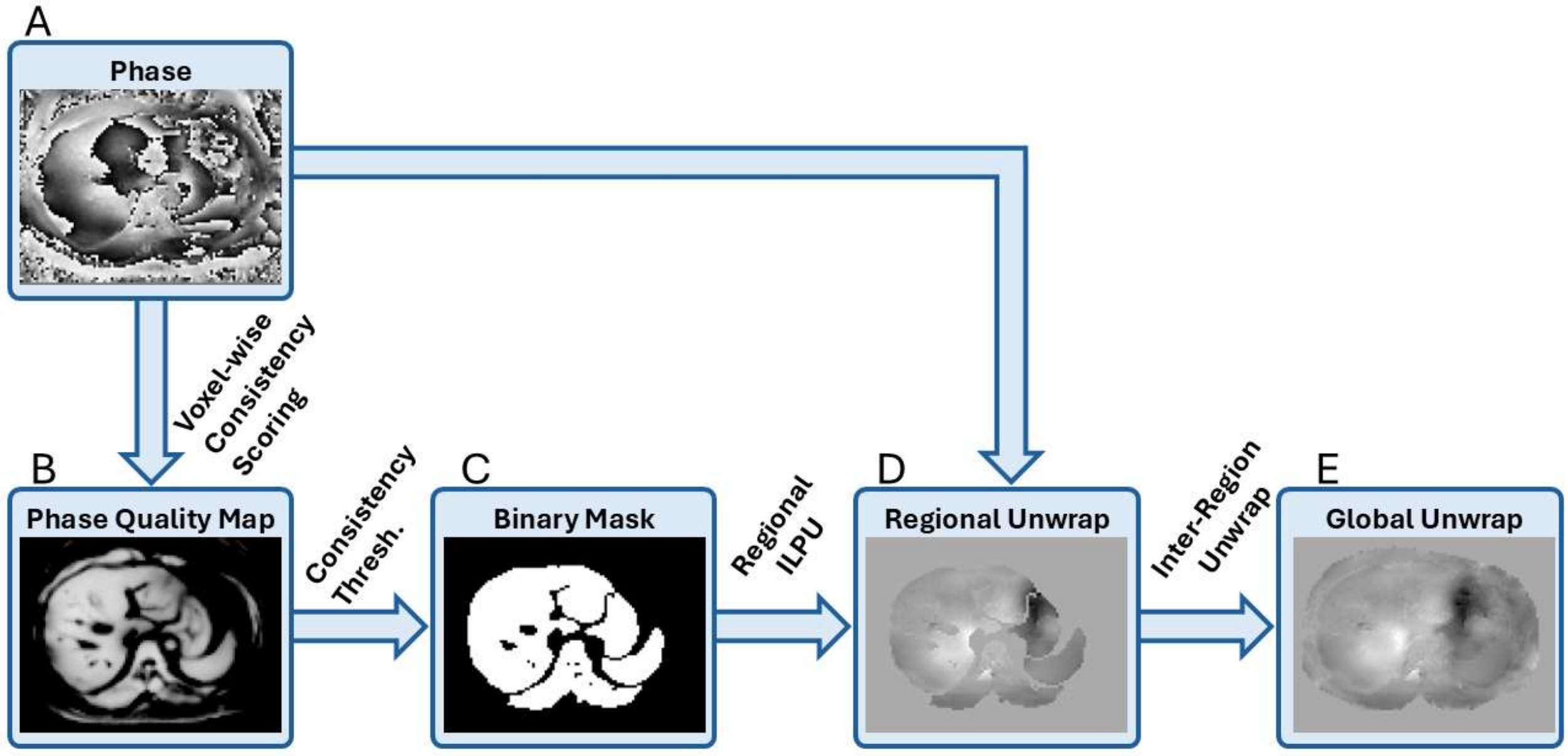

**Figure 1: Spatiotemporal Phase Unwrapping.** (A) Wrapped motion-encoded MRE phase image. (B) Voxel-wise phase-quality map obtained by consistency scoring. (C) Binary reliable-phase mask generated by thresholding the phase-quality map and combining it with the organ ROIs. (D) Regional phase image obtained by applying ILPU separately within each connected organ region. (E) Globally unwrapped phase image obtained by recovering the remaining abdominal voxels using Laplacian phase unwrapping with Dirichlet boundary conditions and the regional ILPU solutions as fixed boundary values.

The shear wavefields were inverted with frequency-resolved k-MDEV [45] to yield maps of shear wave speed (SWS, c, m/s), penetration rate (PR, a, m/s), and loss angle (ϕ, rad) at each drive frequency. The loss angle was derived from SWS and PR by $\phi = 2 \cdot \arctan(c/2a)$. Frequency-resolved maps were used for dispersion analysis and rheological model fitting. For comparison with conventional narrow-band MRE studies, multifrequency k-MDEV was applied within predefined low- (5-20 Hz), mid- (25-40 Hz), and high-frequency (50-80 Hz) bands.

**Phantom Validation**

The accuracy of the wideband acquisition and inversion framework was evaluated using a previously characterized liquid-liver phantom that reproduces the viscoelastic dispersion of healthy human liver tissue [18]. The phantom was scanned from 5 to 80 Hz using the acquisition parameters described above, while mechanical excitation was applied through a vertical plate coupled to a pressurized-air actuator [18]. To quantify the effect of the low-frequency processing adaptations, the wavefields were reconstructed using (i) the default abdominal k-MDEV pipeline and (ii) the proposed wideband pipeline. Frequency-resolved SWS and PR were compared with previously published reference measurements obtained by rheometry, ultrasound elastography, and MRE [18]. Agreement was assessed using Bland-Altman mean bias and 95% limits of agreement; bias, defined as method minus reference, and RMSE were additionally calculated within the predefined low-, mid-, and high-frequency bands.

**Viscoelastic Model Identification**

For each subject, the organ-averaged frequency-resolved SWS and PR were fitted simultaneously (Eq. 1) with six established rheological models. Each model specifies the complex shear modulus $G^*(\omega) = G'(\omega) + iG''(\omega)$ as a function of angular frequency $\omega = 2\pi f$, where $G'$ is the storage modulus and $G''$ the loss modulus. For brevity, the explicit frequency dependence was omitted in the following equations. The following relation was used to derive $G^*$ from SWS and PR:

$$G^* = \rho \left( \frac{1}{\frac{1}{SWS} - i\frac{1}{2PR}} \right)^2. \tag{1}$$

Three two-parameter models (springpot, Kelvin-Voigt, Maxwell) and three three-parameter models (Zener, springpot with parallel spring, springpot with serial Newtonian dashpot) [46] were considered, spanning elastic, viscous, and fractional dissipation behavior. The model functions and free parameters are outlined in Table 2. For each model, the predicted SWS, PR, and loss angle were derived from $G^*$ assuming a homogeneous, isotropic, medium of density $\rho$=1000 kg/m$^3$:

$$SWS = \sqrt{\frac{2|G^*|^2}{\rho(|G^*| + G')}}, \qquad PR = \sqrt{\frac{2|G^*|^2}{\rho(|G^*| - G')}}, \qquad \phi = \arctan\left(\frac{G''}{G'}\right). \tag{2}$$

**Table 2: Overview of viscoelastic models.** Summary of the mechanical elements, the fitted parameters, and the model functions for springpot, Kelvin-Voigt, Maxwell, Zener, springpot with parallel spring, and springpot with serial Newtonian dashpot. Here, || and - denote a parallel and serial element placement, respectively, further $\omega$ denotes angular frequency, $\mu$ elastic modulus, $\eta$ dashpot viscosity, $\alpha$ the springpot power-law exponent. For the springpot elements a specific viscosity of $\eta_0$=7.3 Pa·s for liver tissue was used [18].

| Name | Elements | Parameters | Model |
|---|---|---|---|
| **Springpot** | fractional element | $\mu, \alpha$ | $G^*(\omega) = \mu^{1-\alpha}(i\omega\eta_0)^\alpha$ |
| **Kelvin-Voigt** | spring || dashpot | $\mu, \eta$ | $G^*(\omega) = \mu + i\omega\eta$ |
| **Maxwell** | spring - dashpot | $\mu, \eta$ | $G^*(\omega) = \frac{i\omega\eta\cdot\mu}{\mu+i\omega\eta}$ |

| Zener | spring || (spring - dashpot) | $\mu_1, \mu_2, \eta$ | $G^*(\omega) = \frac{\mu_1\mu_2 + i\omega\eta(\mu_1+\mu_2)}{\mu_2 + i\omega\eta}$ |
|---|---|---|---|
| Springpot with parallel spring | springpot || spring | $\mu_1, \mu_2, \alpha$ | $G^*(\omega) = \mu_1 + \mu_2^{1-\alpha}(i\omega\eta_0)^\alpha$ |
| Springpot with serial dashpot | springpot - dashpot | $\mu, \eta, \alpha$ | $G^*(\omega) = \frac{\mu^{1-\alpha}(i\omega\eta_0)^\alpha \cdot i\omega\eta}{\mu^{1-\alpha}(i\omega\eta_0)^\alpha + i\omega\eta}$ |

Parameters were estimated per subject and organ by nonlinear least-squares optimization under physically admissible bounds (positive modulus and viscosity; $\alpha \in [0, 1]$). Model performance was quantified per subject, organ, and model by the coefficient of determination ($R^2$) and the root-mean-square error (RMSE).

### Statistical Analysis

Analyses were performed in MATLAB R2025a (The MathWorks, Natick, MA). Variables are summarized as median [interquartile range, IQR]. Group differences between healthy volunteers and patients were assessed for each organ, frequency-band and fitted model parameter with two-sided Wilcoxon rank-sum tests. Significance was set at $p<0.05$.

## Results

### Liquid-Liver Reference Phantom

When the phantom data were processed with the default abdominal k-MDEV inversion, SWS was systematically underestimated at low frequencies relative to both the reference data and the wideband inversion, with the two inversions converging only above 25 Hz (see Supplemental Note S1). This low-frequency limitation of the default pipeline motivated the frequency-adaptive wideband inversion evaluated here. Agreement between the proposed wideband MRE framework and previously published phantom reference data [18] is demonstrated by the Bland-Altman analysis shown in Figure 2. Only a small positive bias for SWS and PR measured by ultra-wideband MRE relative to the reference dataset was observed. For SWS, the mean bias was 0.035 m/s, with a standard deviation of differences of 0.046 m/s and 95% limits of agreement (LoA) from -0.056 to 0.126 m/s. For PR, the mean bias was 0.016 m/s, with a standard deviation of differences of 0.073 m/s and 95% LoA from -0.127 to 0.159 m/s.

The dispersion curves in Figure 2 illustrate the agreement between wideband MRE and reference measurements across the investigated frequency range. For both SWS and PR, wideband MRE reproduced the expected frequency-dependent increase observed in the previously published data.

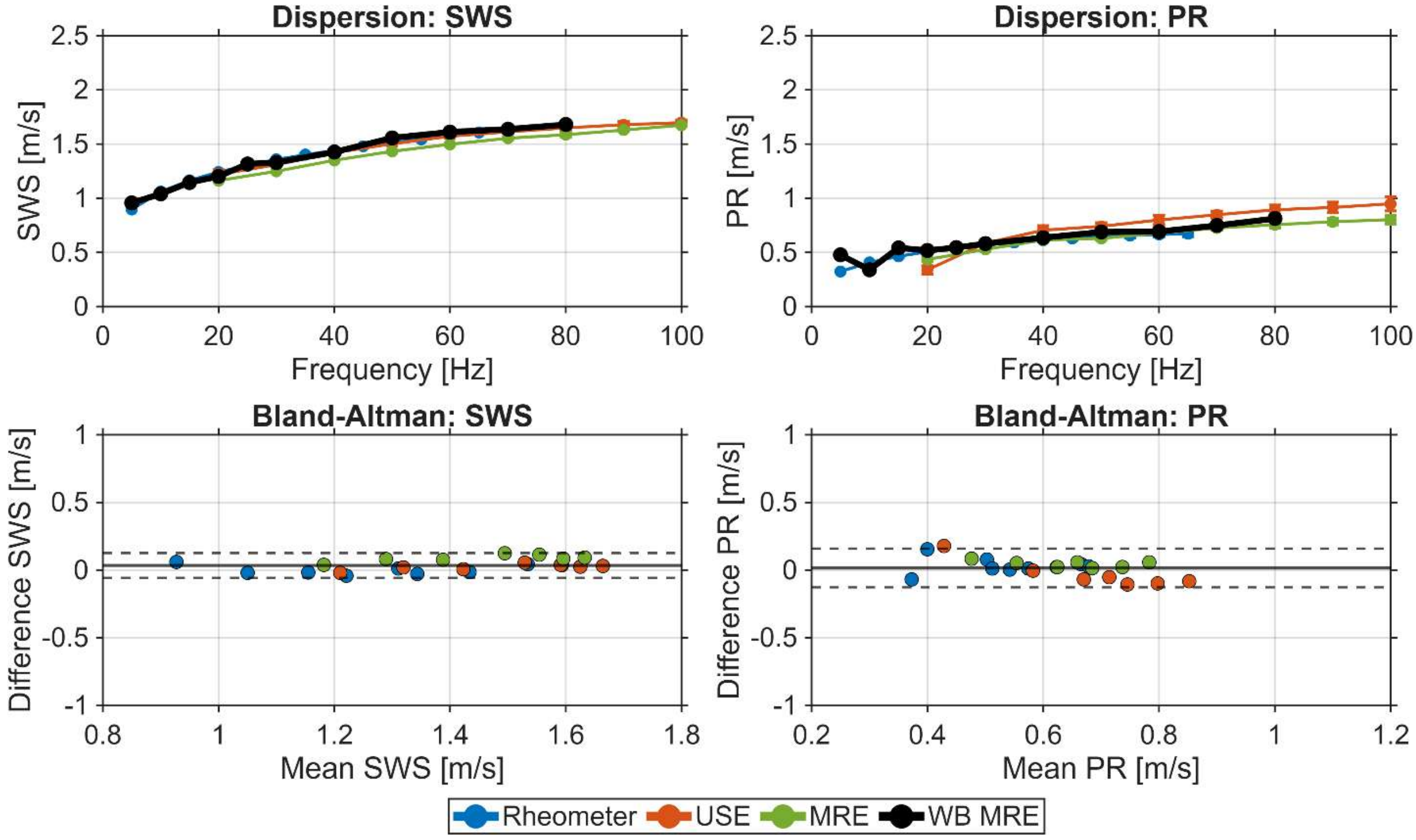

**Figure 2: Dispersion and agreement analysis of the liquid-liver phantom.** (Top row) Frequency-resolved shear wave speed (SWS) and penetration rate (PR) measured and processed with the proposed wideband MRE framework (black) are compared with previously published rheometer (blue), ultrasound elastography (USE, orange) and MRE (green) reference data. (Bottom row) Bland-Altman analysis for SWS and PR comparing wideband MRE with the matched reference measurements. Points are color-coded according to the reference acquisition type. The solid horizontal line indicates the mean bias, and dashed horizontal lines indicate the 95% LoA.

### Shear Wavefields and Viscoelastic Parameter Maps

Representative wavefield and parameter maps are shown in Figure 3. The shear wavefield propagated through both the liver and spleen. The maps of SWS, PR, and ϕ captured complementary mechanical information: SWS reflected the stiffness-related component, PR the spatial wave damping, and ϕ the balance between elastic energy storage and viscous energy loss.

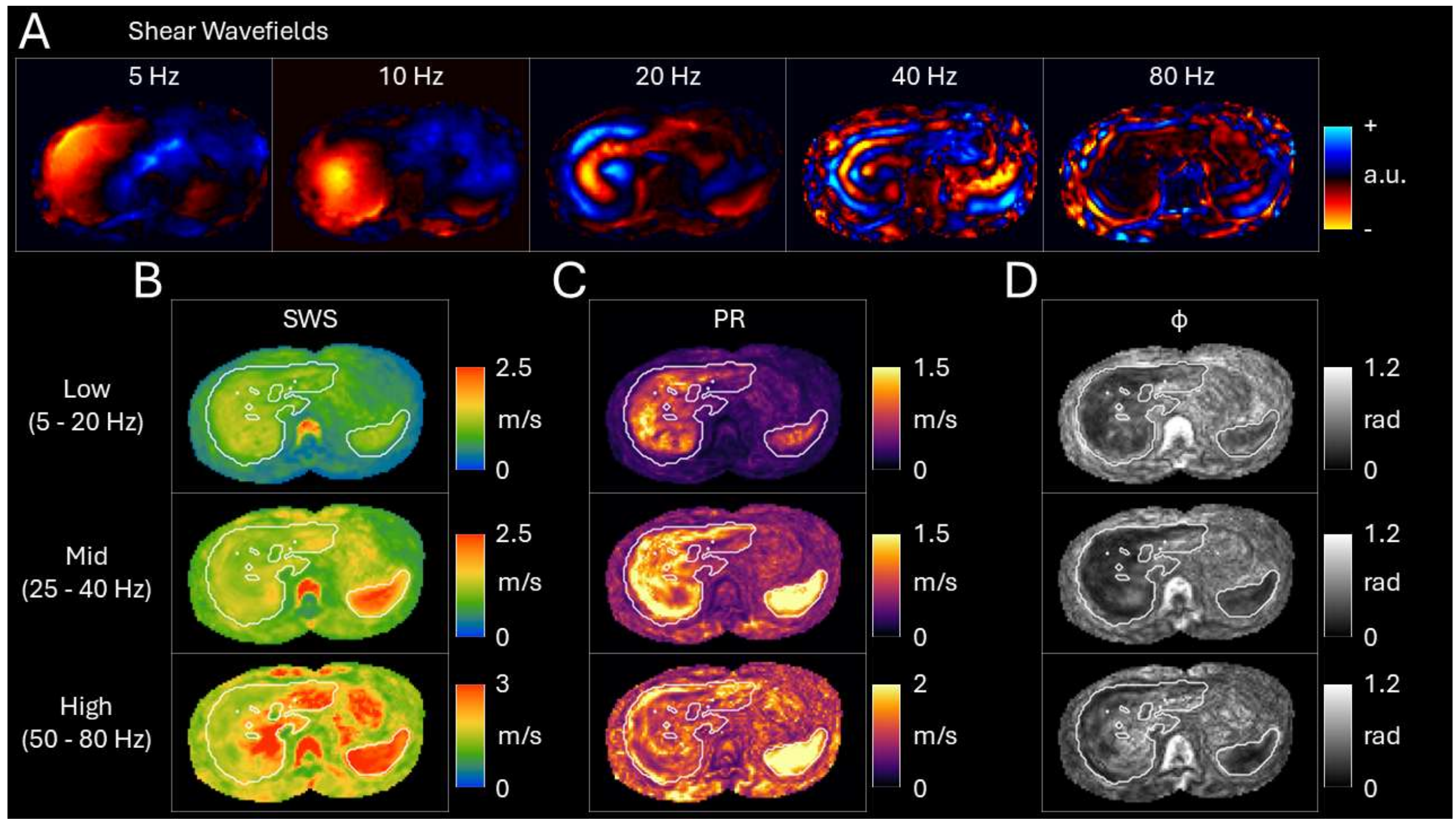

**Figure 3: Representative shear wavefield, SWS, PR, and ϕ maps for a healthy volunteer.** (A) Shear wavefields at 5, 10, 20, 40, and 80 Hz vibration frequency. (B-D) Maps of shear wave speed (SWS), penetration rate (PR), and loss angle (ϕ) processed through multifrequency k-MDEV inversion in 5-20 Hz, 25-40 Hz, 50-80 Hz. The white outline indicates the liver and spleen ROIs.

### Frequency-Dependent Viscoelastic Parameters

To assess the sensitivity of MRE at different frequency ranges, k-MDEV inversion was performed in ranges of 5-20 Hz, 25-40 Hz and 50-80 Hz drive frequency. Both organs showed a pronounced viscoelastic dispersion (Figure 4). In healthy volunteers, median liver SWS increased from 1.05 m/s in the low-frequency band to 1.61 m/s in the high-frequency band, while spleen SWS increased from 1.36 m/s to 2.11 m/s. Similarly, PR exhibited a marked frequency dispersion (liver: 0.59 to 1.45 m/s; spleen: 0.43 to 2.14 m/s). In contrast, loss angle decreased from 0.60 to 0.34 rad in the liver and from 0.97 to 0.31 rad in the spleen.

Strongest group differences between healthy volunteers and patients were observed primarily in the liver within the low-frequency regime. Patients showed a higher ϕ (median [IQR]: 0.99 [0.80, 1.04] rad) and 36.6% lower PR (0.37 [0.33, 0.50] m/s) than healthy volunteers (ϕ: 0.60 [0.59, 0.69] rad, $p<0.001$, PR: 0.59 [0.53, 0.64] m/s, $p=0.003$). Liver SWS was only modestly increased about 7.7% in patients (1.13 [1.08, 1.16]) compared to healthy volunteers (1.05 [1.03, 1.06] m/s, $p=0.008$). In the spleen, no parameter differed significantly between healthy volunteers and patients.

A table with median and IQR values for MRE parameters in the different frequency ranges of the liver and spleen is provided in Supplemental Table S3.

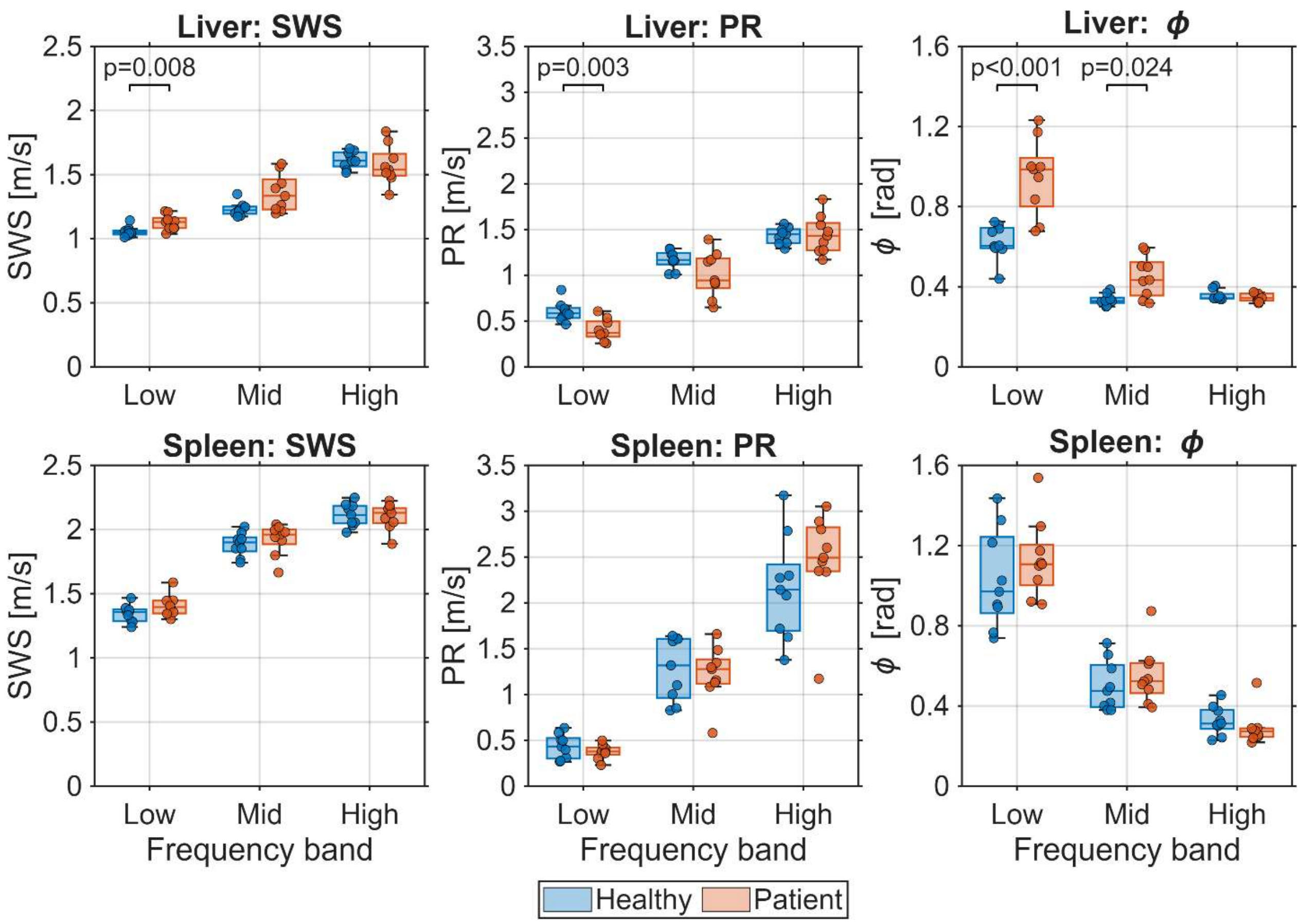

**Figure 4: Frequency-band averaged viscoelastic parameters in healthy volunteers and patients.** Shear wave speed (SWS), penetration rate (PR), and loss angle (ϕ) of liver (top) and spleen (bottom), derived from multifrequency inversion within low (5-20Hz), mid (25-40 Hz), and high (50-80 Hz) frequency bands. Healthy volunteers (blue) and patients (red) are compared with Wilcoxon rank-sum tests. Patients showed an increased low-frequency liver ϕ and reduced PR, whereas high-frequency and splenic differences were not significant.

**Frequency-Resolved Viscoelastic Behavior of the Liver and Spleen**

Figure 5 shows the MRE measured viscoelastic properties of the liver and spleen from 5 to 80 Hz along with the fitted SP-dashpot model for healthy volunteers and patient data respectively. Both organs exhibited pronounced dispersion in SWS, PR, and ϕ across the investigated frequency range. In the liver, SWS increased progressively with excitation frequency, indicating a gradual increase in stiffness of the tissue at higher frequencies. Similarly, PR increased with frequency, demonstrating increased wave attenuation and shorter propagation distances for higher-frequency shear waves. In contrast, the loss angle showed a monotonic decrease across the investigated frequency range, with the highest values observed at low frequencies and progressively lower values at higher frequencies. Fitted $G'$, $G''$, and $|G^*|$ dispersion curves are shown in Supplemental Figure S4.

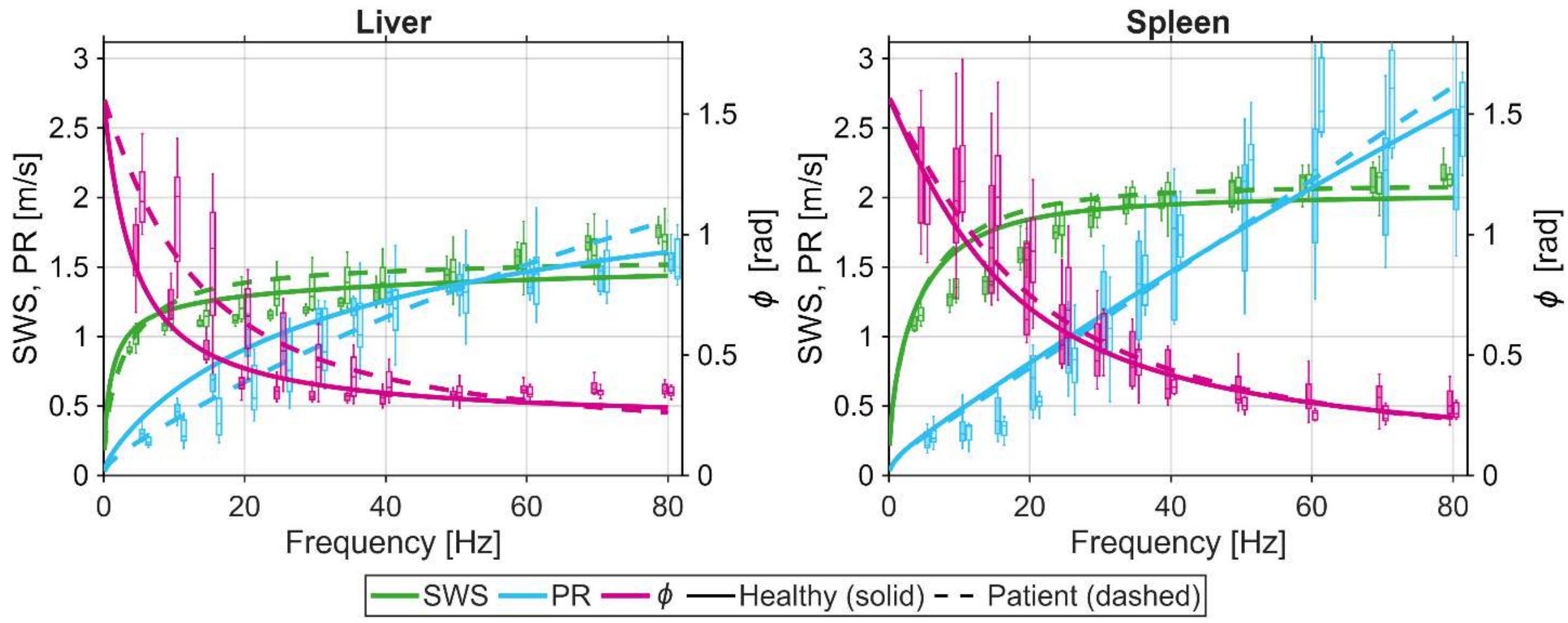

**Figure 5: Measured and model-fitted viscoelastic dispersion of the liver and spleen.** Measured data and fitted SP-dashpot dispersion curves are shown for the liver (left) and spleen (right) across the investigated frequency range. Boxplots show the MRE measured values while lines indicated model fits. Healthy volunteers are shown with solid lines and patients with dashed lines.

The relative differences in MRE parameters between liver and spleen were highly frequency-dependent (Figure 6). SWS was consistently higher in the spleen than the liver across all frequencies. However, SWS contrast was lowest in the low-frequency band (21.0% [18.0, 24.3]), with a maximum of 50.5% [44.0, 55.0] in the mid-frequency band, and a reduction again to 36.3% [25.5, 45.2] in the high-frequency band. The PR contrast was negative at low frequencies (-24.2% [-50.4, -12.4]), negligible in the mid-frequency band (-4.0% [-24.5, 16.9]), and substantially higher at high frequencies (+43.9% [19.2, 63.0]). The ϕ contrast showed the opposite tendency with mean differences between spleen and liver of 17.8% [4.4, 32.6], 24.5% [12.6, 36.7] and -2.5% [-15.3, 9.1] from low, to mid, to high-frequency ranges.

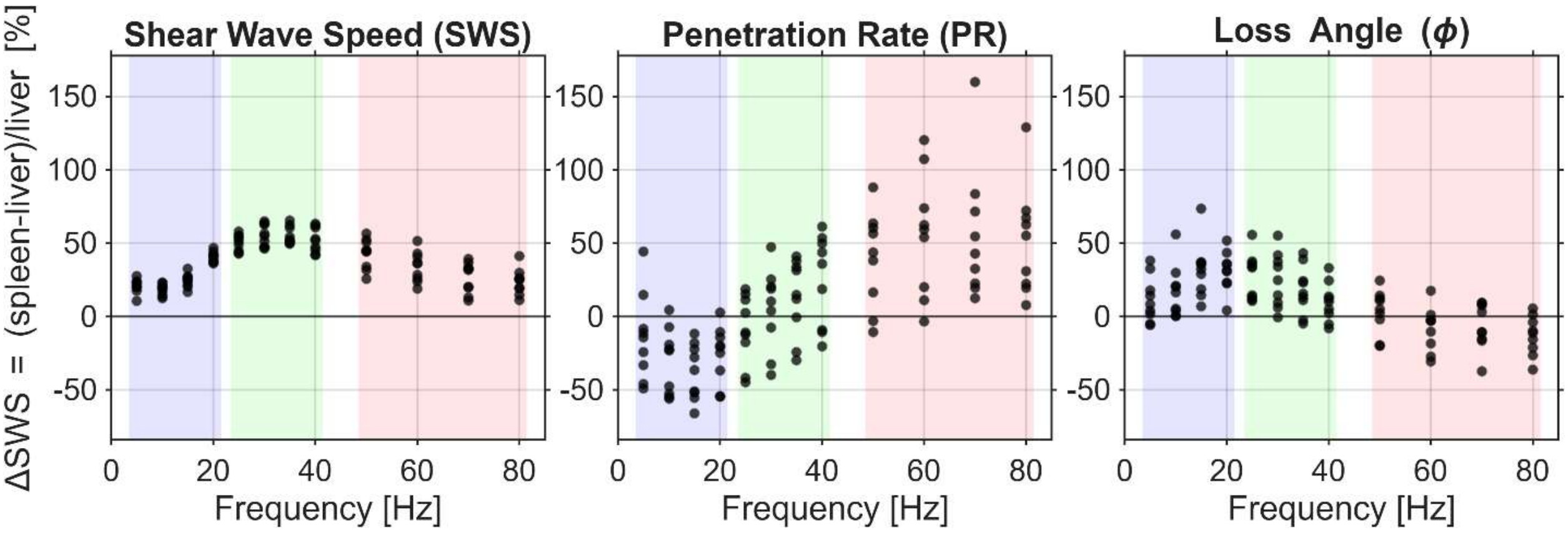

**Figure 6: Relative difference between frequency-resolved viscoelastic parameters of the liver and spleen.** The relative difference of SWS, PR, and ϕ of the liver and spleen are plotted over frequency for healthy volunteers. The colored background indicates the low (blue), mid (green), and high (red) frequency ranges.

### Viscoelastic Model Identification

The six investigated rheological models showed distinct frequency-dependent relationships between SWS, PR, and ϕ. Mean fitted dispersion curves for all models for the liver are shown in Figure 7. The different dispersion properties provided insights into the ability of each model to reproduce the experimentally observed viscoelastic dispersion behavior of the liver and spleen.

The springpot model predicted power-law increases in both SWS and PR with increasing frequency while maintaining a constant ϕ across the entire frequency range. The springpot with parallel spring converged to a springpot, omitting any influence of the parallel spring as discussed in [13].

In contrast, the Kelvin-Voigt model predicted increasing SWS and ϕ with frequency, while PR decreased. This behavior differs substantially from the experimentally observed trends, where PR increased and ϕ decreased with frequency (Figure 5).

The Maxwell model generated increasing SWS and PR accompanied by a decreasing ϕ. This qualitative behavior was more consistent with the measured dispersion curves of Figure 5. Similarly, the Zener model predicted an increase in both SWS and PR, alongside a decrease in ϕ at higher frequencies. However, at frequencies below 10 Hz, PR increased while ϕ decreased to zero. Neither of these low-frequency effects represented the measured data.

Among all investigated models, the SP-dashpot model most closely reproduced the characteristic dispersion behavior observed experimentally. The model-predicted a simultaneous increase in SWS and PR with frequency while exhibiting a monotonic decrease in ϕ. These trends closely resembled the measured dispersion characteristics of both the liver and spleen.

Substantial differences in fitting performance were observed among the tested models. In the liver, the SP-dashpot model achieved the highest goodness-of-fit, yielding a median $R^2$ of 0.88 [0.85, 0.92] and the lowest RMSE of 0.105 [0.094, 0.112]. The Zener model provided the second-best performance with an $R^2$ of 0.51 [0.34, 0.82] and an RMSE of 0.207 [0.159, 0.233], followed by the springpot and springpot with parallel spring models, which showed the same fitting accuracy. The Maxwell model demonstrated only moderate agreement with the measured data, whereas the Kelvin-Voigt model failed to describe the observed liver dispersion behavior, resulting in strongly negative $R^2$ values. In the spleen, the SP-dashpot model achieved the highest median $R^2$ of 0.91 [0.87, 0.92], but the Maxwell (0.89 [0.83, 0.92]) and Zener (0.90 [0.84, 0.93]) performed similarly, indicating that splenic dispersion did not uniquely favor the SP-dashpot model. An overview of the $G^*$-based goodness-of-fit values is provided for each model in Table 3. The measured and model-predicted dispersion curves for $G'$, $G''$, and $|G^*|$ are provided in Supplemental Figures S2 and S3.

Although the SP-dashpot contains an additional parameter and considers the springpot as the limiting case $\eta \rightarrow \infty$, improvements in both $R^2$ and RMSE suggest that the serial dashpot captures dispersion behavior not represented by a springpot alone. The fitted values of η are also consistent with a contribution from the serial dashpot rather than convergence to the springpot limit.

As the SP-dashpot model provided the best goodness-of-fit, the model was selected for subsequent analyses of viscoelastic model parameters between healthy volunteers and patients. The measured data as well as the fitted SP-dashpot dispersion curves are shown in Figure 5.

In the liver, patients showed significant alterations in all three model parameters. The power-law exponent $\alpha$ was significantly lower in patients than in healthy volunteers (0.06 [0.03, 0.09] compared to 0.13 [0.11, 0.15]; $p<0.001$). Similarly, the dashpot viscosity $\eta$ was reduced in patients (23.1 [20.1, 38.1] Pa·s compared to 48.7 [38.6, 58.9] Pa·s; p=0.002). In contrast, the springpot elastic modulus $\mu_{SP}$ was significantly elevated in patients (1.82 [1.16, 1.95] kPa compared to 0.90 [0.84, 1.02] kPa; p=0.001). In the spleen, only $\mu_{SP}$ differed significantly, being higher in patients than in healthy volunteers (4.15 [3.91, 4.65] kPa compared to 3.43 [3.08, 3.75] kPa; p=0.003). The dashpot viscosity $\eta$ and power-law exponent $\alpha$ showed similar trends to those observed in the liver but were not significant.

**Table 3: Comparison of rheological model performance in liver and spleen**. Goodness-of-fit metrics for the fitted rheological models are reported as median [IQR]. Model performance was assessed using the coefficient of determination ($R^2$) and root-mean-square error (RMSE) between $G^*$ of the model and the acquired data. Higher $R^2$ and lower RMSE indicate improved agreement between the measured and model-predicted dispersion curves. The SP-dashpot model provided the best goodness-of-fit for both the liver and spleen (row marked in green).

| | Liver | | Spleen | |
|---|---|---|---|---|
| | $R^2$ | RMSE | $R^2$ | RMSE |
| **Springpot** | 0.46 [0.38, 0.48] | 0.22 [0.20, 0.29] | 0.29 [0.24, 0.35] | 0.38 [0.33, 0.43] |
| **Kelvin-Voigt** | -4.48 [-6.13, -2.07] | 0.73 [0.64, 0.76] | -1.22 [-1.87, -0.88] | 0.70 [0.66, 0.73] |
| **Maxwell** | 0.33 [0.04, 0.81] | 0.25 [0.16, 0.28] | 0.89 [0.83, 0.92] | 0.15 [0.13, 0.18] |
| **Zener** | 0.51 [0.34, 0.82] | 0.21 [0.16, 0.23] | 0.90 [0.84, 0.93] | 0.15 [0.12, 0.17] |
| **SP \|\| spring** | 0.46 [0.38, 0.48] | 0.22 [0.20, 0.29] | 0.29 [0.24, 0.35] | 0.38 [0.33, 0.43] |
| **SP-dashpot** | **0.88 [0.85, 0.92]** | **0.11 [0.09, 0.11]** | **0.91 [0.87, 0.92]** | **0.14 [0.12, 0.15]** |

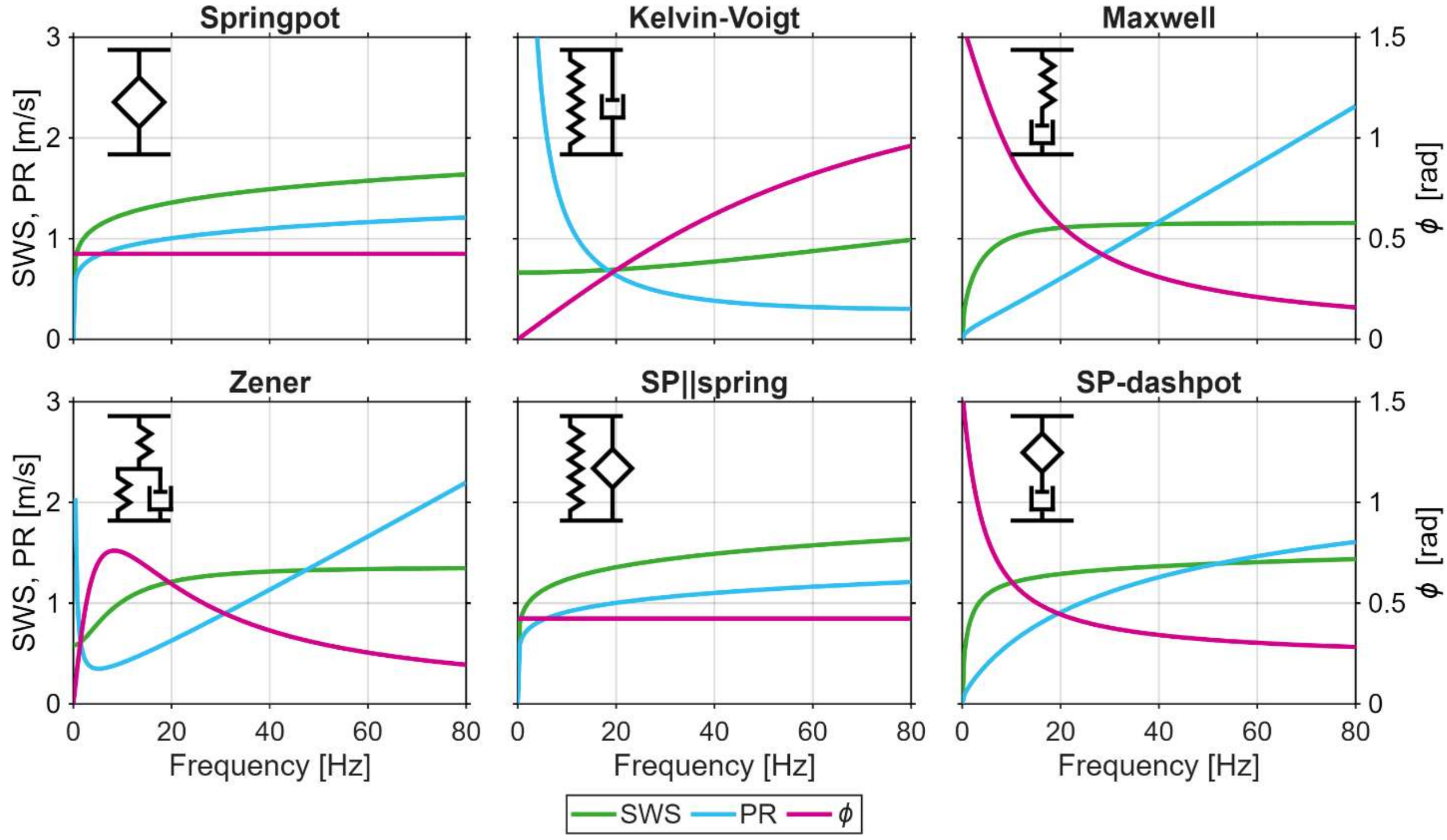

**Figure 7: Model-predicted dispersion behavior of liver tissue.** Visualization of the frequency dependence of cohort-averaged dispersion curves for the six viscoelastic models. Shear wave speed (SWS, green) and penetration rate (PR, blue) are shown on the left y-axis and loss angle (ϕ, magenta) is shown on the right y-axis.

**Effect Sizes and Diagnostic Performance**

To compare the effect sizes and diagnostic performance of the observed viscoelastic alterations, the relative differences between healthy volunteers and patients were assessed for the most significant parameters: the unmodelled low-frequency ϕ and the full-frequency range fitted SP-dashpot model parameters of the liver. Figure 8 summarizes the relative changes in these four parameters.

Among the measured mechanical parameters, low-frequency ϕ showed a 63.2% increase ($p<0.001$) from 0.60 rad in healthy volunteers to 0.99 rad in patients, which was accompanied by changes in all three fitted model parameters. The power-law exponent α showed the largest relative reduction, decreasing by 58.0% ($p<0.001$) from 0.13 in healthy volunteers to 0.06 in patients. Similarly, the dashpot viscosity η decreased by 52.5% ($p=0.002$), from 48.7 Pa·s to 23.1 Pa·s. In contrast, the springpot elastic modulus $\mu_{SP}$ increased by approximately 101% ($p=0.001$) from 0.90 kPa in healthy volunteers to 1.82 kPa in patients.

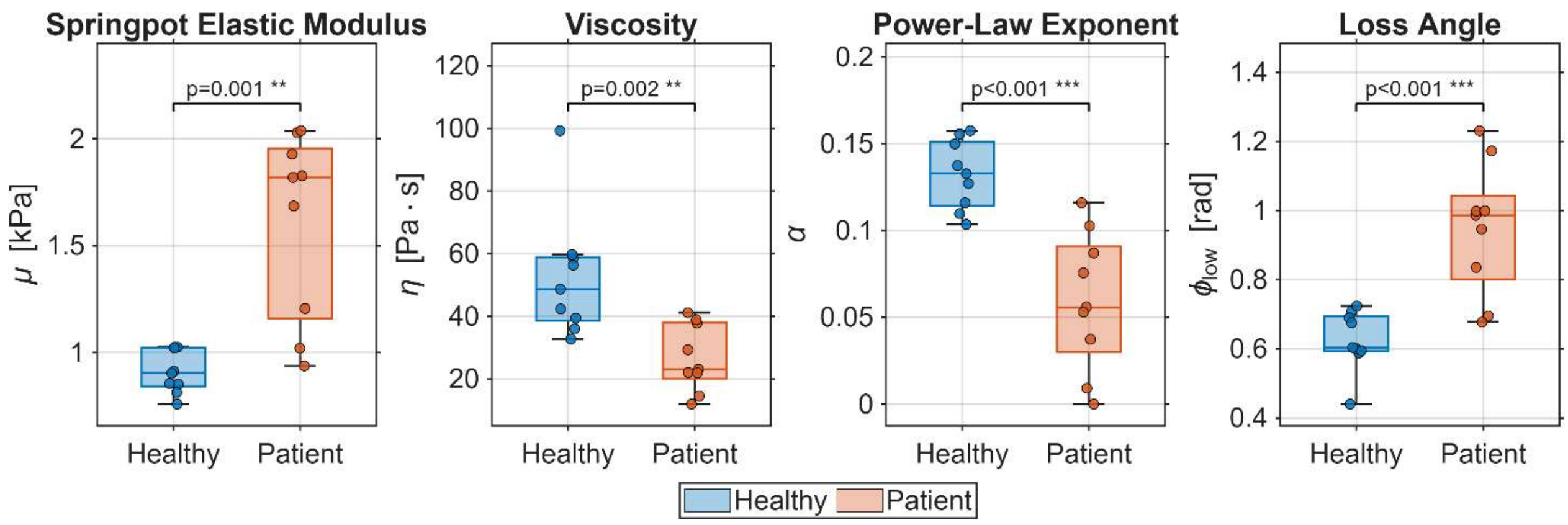

**Figure 8: SP-dashpot model parameters and Loss Angle of the liver in healthy volunteers and patients.** Full-frequency range (5-80 Hz) fitted values of the springpot elastic modulus μ, dashpot viscosity η, and power-law exponent α from wideband liver dispersion data according to the SP-dashpot model. Additionally, measured and unmodelled loss angle $\phi_{low}$ is shown for the low-frequency range of 5-20 Hz. Data from healthy volunteers are shown in blue, patients in red. Wilcoxon rank-sum p-values are shown as indication of significance.

To further specify the frequency dependence of diagnostic power, liver ϕ was analyzed at each individual excitation frequency (Figure 9A). Consistent with the band-based analysis, patients showed higher ϕ values than healthy volunteers predominantly at low frequencies, with the largest relative differences observed between 5 and 30 Hz. The percent difference increased from 5 Hz to a maximum around 15 Hz and subsequently declined with increasing frequency (Figure 9B). Frequency-resolved AUROC analysis (Figure 9C) showed highest discriminatory performance (>0.80) across the low-frequency range, whereas diagnostic separation progressively decreased at higher frequencies and was negligible within the conventional clinical MRE range (>40 Hz). These results indicate that the significance of $\phi_{low}$ was driven by a robust inflammation-associated increase in liver ϕ with peak effect sizes around 15 Hz, suggesting this frequency range for MRE in hepatic inflammation.

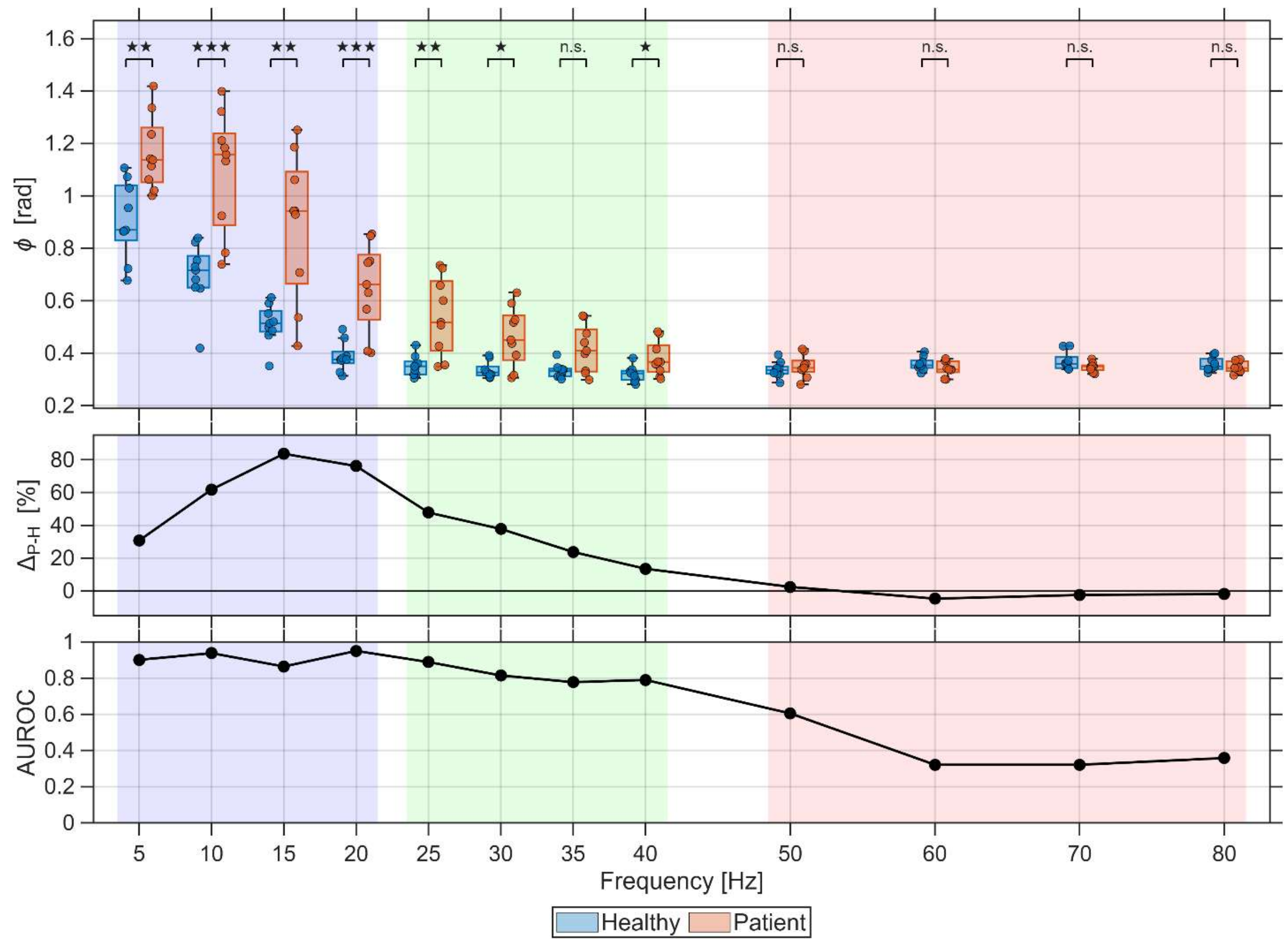

**Figure 9: Frequency-resolved diagnostic performance of loss angle**. A) Liver loss angle (ϕ) at individual excitation frequencies from 5 to 80 Hz in healthy volunteers and patients. Significance was assessed using two-sided Wilcoxon rank-sum tests. B) Relative percent difference in ϕ between patients and healthy volunteers, calculated for each frequency. C) Frequency-resolved AUROC for discrimination between patients and healthy volunteers based on liver ϕ. Diagnostic performance was highest at low frequencies and progressively decreased within the conventional clinical MRE frequency range. Shaded regions indicate low- (blue), mid- (green), and high- (red) frequency ranges.

## Discussion

This study demonstrates the feasibility of ultra-wideband MRE of the human liver and spleen across an unprecedentedly wide frequency range of 5 to 80 Hz and shows that frequency-resolved viscoelastic parameters are sensitive to hepatic inflammation. The most pronounced patient-related differences were observed at frequencies below 20 Hz, with a 63% increase in loss angle and a 37% reduction in penetration rate, accompanied by only a modest 8% increase in shear wave speed. Together, these findings suggest that hepatic inflammation primarily alters the dissipative, fluid-like mechanical behavior rather than tissue stiffness. Rheological modeling linked the measured dispersion to underlying material properties. Among the investigated models, the SP-dashpot provided the best description of the measured dispersion.

MRE has become an established tool for the noninvasive assessment and staging of liver fibrosis. Current clinical implementations, however, are typically performed within a narrow frequency range of approximately 40-60 Hz and generally report only a single tissue stiffness value [47]. While this approach is well suited for characterizing fibrosis, hepatic inflammation remains difficult to assess noninvasively.

Achieving reliable measurements across the ultra-wideband frequency range required several methodological developments. At low excitation frequencies, shear wavelengths become comparable to liver dimensions making reconstruction particularly challenging. For example, assuming a shear modulus of 100 Pa at 1 Hz [14] yields wavelengths on the order of 32 cm, whereas a shear modulus of approximately 800 Pa at 5 Hz results in wavelengths of about 18 cm, comparable to those in cirrhotic livers at conventional clinical frequencies. Accurate recovery of these relatively long wavelengths required a combination of phase unwrapping such as ILPU that does not impose smoothness constraints across organ boundaries [39], frequency-adaptive bandpass filtering [16], and noise-robust wavenumber-based inversion [44]. Together, these developments enabled stable reconstruction across a wide range of excitation frequencies despite substantial differences in wavelength and displacement amplitude.

The reliability of the proposed method was validated by phantom experiments, which reproduced the expected frequency dependence of SWS and PR with only small deviations from previously reported rheometry, ultrasound elastography, and MRE reference measurements [18].

In healthy volunteers, both the liver and spleen showed pronounced frequency-dependent viscoelastic behavior across the investigated range. While the spleen remained consistently stiffer than the liver, the relative mechanical contrast between the two organs varied with excitation frequency. Within the clinical MRE frequency range, the spleen is much stiffer than the liver, however, the relative difference decreases toward lower frequencies. During manual palpation, the spleen is generally perceived as softer than the liver. Palpation is a quasi-static, large-strain assessment of bulk organ compliance, influenced by boundary conditions and slow redistribution of blood in the vascular spleen. MRE, in contrast, probes the dynamic shear response at vibration frequencies, at which the dense splenic microstructure resists shear more than hepatic parenchyma. Higher splenic than hepatic stiffness has been reported in previous MRE studies [35], indicating that the spleen's stiffer dynamic behavior and softer palpatory feel reflect different deformation modes and the pronounced poroelastic behavior of splenic tissue. In our cohort, the spleen provided limited discriminatory information between the healthy volunteers and patients with hepatic inflammation. The observed splenic alterations may primarily reflect systemic or hemodynamic consequences of liver disease rather than inflammatory changes within the splenic parenchyma [48,49]. Consequently, their diagnostic value should be investigated in cohorts with portal hypertension or advanced chronic liver disease.

Our observation of marked low-frequency dispersion of liver stiffness is supported by recent studies , which exploited intrinsic, cardiovascular-pulsation induced hepatic strain at a frequency of approximately 1 Hz [14,50–53]. Baradaran Najar et al. combined intrinsic MRE with conventional extrinsic MRE at 30-60 Hz. They reported an increase in the magnitude of the complex shear modulus of the liver by a factor of approximately 25.5 from 1 Hz to 60 Hz, consistent with the SWS dispersion we found on a continuous frequency axis. By resolving viscoelastic parameters continuously from 5 to 80 Hz, our work spans four octaves of excitation frequency and bridges these intrinsic physiological excitation frequencies with the established clinical MRE range. This pronounced low-frequency dispersion mirrors the strong dispersion previously reported for intrinsic MRE of the brain [16].

Rheological modeling across the full frequency range provided a complementary interpretation of the measured frequency dispersion by constitutive material properties. Among the six candidate models, the SP-dashpot provided the best description of both liver and spleen dispersion while simpler models with only two parameters could not reproduce low-frequency dispersion. The increased springpot elastic modulus together with the reduced dashpot viscosity and power-law exponent, indicated an inflammation-associated property shift toward a stronger dissipation and fluid-like behavior at low frequencies in the liver. The observed reduction in the dashpot power-law exponent is consistent with previous multifrequency MRE studies of hepatic inflammation [54]. Baradaran Najar et al. reported that a multifrequency dispersion coefficient derived from 30, 40, and 60 Hz MRE decreased with increasing lobular inflammation and remained independently associated with inflammation after adjustment for steatosis, ballooning, and fibrosis [37]. Similarly, Bayerl et al. reported high diagnostic performance of an MRE-derived dispersion parameter, again in the range of 30 to 60 Hz, for grading hepatic inflammation [38]. The present results agree with these findings by showing a marked reduction in the fitted power-law exponent in patients, while extending the analysis to a substantially wider frequency range. Importantly, our results show that the strongest inflammation-associated changes occur below the conventional clinical frequency range.

Our findings motivate the development of dual-dynamics MRE of the liver that integrates low- and high frequencies to probe complementary pathological processes of chronic liver disease. Based on the observed frequency dependence, the low-frequency component (<20 Hz) could be used to detect changes in tissue fluidity and viscoelastic dissipation associated with inflammation, whereas the standard clinical high-frequency component (>40 Hz) could be used to probe fibrosis-related stiffening. Such an approach may improve the simultaneous characterization of inflammatory activity and fibrosis; two critical processes in the pathological cascade of liver disease.

Although the present implementation used dense temporal sampling to ensure robust characterization of low-frequency wave dynamics, previous work has demonstrated that accurate MRE reconstruction can be achieved with a reduced number of samples per vibration cycle [55]. Future protocol optimization should combine reduced temporal sampling with targeted acquisition of the most diagnostically informative low- and high-frequency ranges. This is expected to limit scan times to less than 3 minutes while preserving the frequency-resolved biomechanical information and making it compatible with routine abdominal MRI workflows.

Several limitations should be considered. The study was performed in a relatively small cohort using a single-center, cross-sectional design because of the extensive exploratory data acquisition protocol of up to 96 acquisitions per slice and frequency. Furthermore, the patient group was defined by clinically diagnosed inflammatory liver disease and likely represents a limited and heterogeneous disease spectrum. Validation in larger, well-characterized clinical cohorts is required for generalization and sub-group analysis. Finally, physiological confounders, including portal pressure, hepatic perfusion, body mass index, and the severity of inflammation, were not controlled and may have influenced the measured biomechanical parameters. Nonetheless, despite these biological variations, the robust effect sizes particularly of loss angle in the low-frequency range observed in this study encourage larger clinical trials to test MRE in hepatic inflammatory disease.

In summary, ultra-wideband MRE enables continuous biomechanical characterization of the liver and spleen across frequencies from 5 to 80 Hz. Hepatic inflammation was associated with pronounced alterations in low-frequency dissipative mechanical behavior that were largely absent within the conventional clinical MRE frequency range. Rheological modeling provided a complementary mechanistic description of these changes through an increased springpot elastic modulus together with reduced dashpot viscosity and springpot power-law exponent. Collectively, these findings establish low-frequency wideband MRE as a promising approach for extending conventional liver MRE from single-frequency stiffness assessment toward comprehensive frequency-resolved characterization of hepatic tissue mechanics.

## Funding

This work was supported by the German Research Foundation [grant numbers 460333672 CRC1540 EBM, RTG2260 BIOQIC, CRC1340 Matrix-in-vision, FOR5628, 540759292 Sa901/33-1 M5]

# Supplemental

**Supplemental Table S1: Clinical diagnoses of patients with inflammatory liver disease.**

| Diagnosis | Count |
|---|---|
| Hepatic Porphyria | 3 |
| Hemochromatosis | 2 |
| Budd-Chiari syndrome | 1 |
| Steatotic liver disease (MASLD / ALD) | 3 |

**Supplemental Note S1: Comparison of the wideband inversion with the default k-MDEV inversion**

The proposed wideband inversion was developed to recover the relatively long shear wavelengths at low excitation frequencies. The default abdominal k-MDEV pipeline, however, was not designed to operate within this regime. To quantify the differences between the two inversions, a wideband phantom dataset (5-80 Hz) was processed using both pipelines and compared with previously published reference data.

The default k-MDEV inversion uses single-step global Laplacian phase unwrapping and a fixed linear high-pass filter to isolate shear waves. In contrast, the wideband inversion uses iterative Laplacian-based phase unwrapping (ILPU) with frequency-adaptive high-pass filtering, in which the cut-off wavenumber scales with driver frequency below 20 Hz. Accuracy was assessed using liquid-liver reference phantom data (rheometry, ultrasound elastography, MRE). The difference between each inversion and the reference is reported as band wise bias and RMSE, see Table S2.

The two inversions behaved very differently for SWS. At and above 25 Hz, the results were equivalent to the reference for both pipelines, with SWS bias ≤0.08 m/s. Below 20 Hz, however, the default abdominal inversion increasingly underestimated SWS, reaching a deviation of 0.44 m/s (49%) at 5 Hz and a band wise bias of -0.28 m/s (RMSE 0.31 m/s) across 5-20 Hz. In contrast, the wideband inversion remained within 0.06 m/s of the reference (band wise bias -0.02 m/s, RMSE 0.04 m/s), see Figure S1 and Table S2.

The PR was less sensitive to the inversion and did not show the severe underestimation at low frequencies. Both inversions agreed with the reference across the full frequency range (PR bias ≤0.09 m/s for ≥25 Hz). In the low-frequency regime, the default inversion deviated by 0.038 m/s compared to 0.053 m/s for the wideband inversion.

These results suggest that the abdominal k-MDEV inversion, which was developed for the clinical abdominal MRE frequency range of 30 to 50 Hz, is not suitable for the long wavelengths encountered below 25 Hz, and that a dedicated wideband inversion is necessary for unbiased low-frequency dispersion mapping.

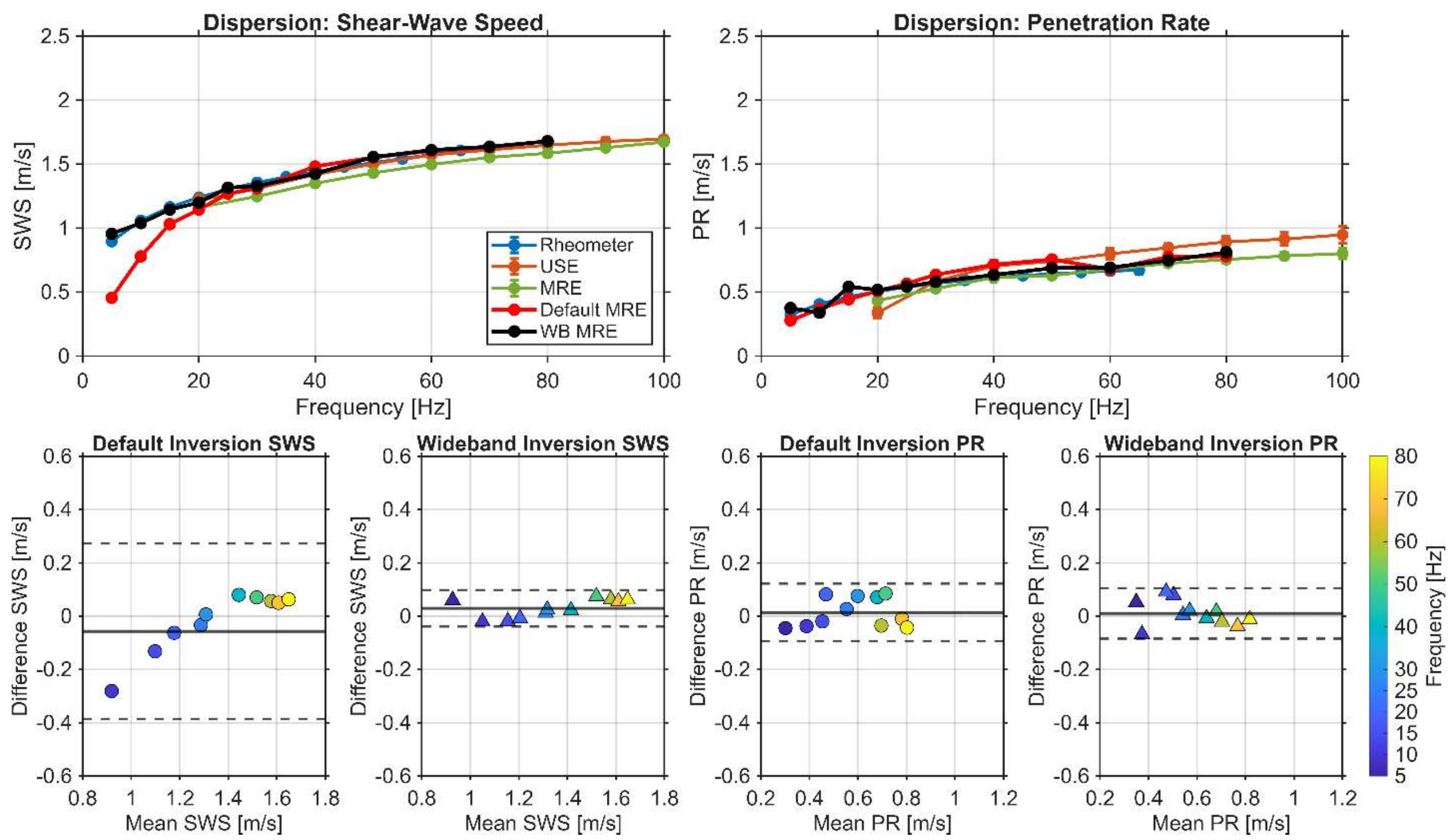

**Supplemental Figure S1:** Comparison of the default k-MDEV and wideband inversions. (A, B) Frequency-resolved SWS and PR of the liquid-liver phantom reconstructed with the default k-MDEV inversion (red) and the wideband inversion (black), overlaid on the independent reference measurements (colored). (C-F) Bland-Altman plots comparing the default and wideband inversion against the matched reference for SWS and PR. Individual data points are colored by frequency.

**Supplemental Table S2:** Accuracy of the two inversions relative to the liquid-liver reference within the low- (5-20 Hz), mid- (25-40 Hz), and high-frequency (50-80 Hz) bands. Bias (method - reference) and RMSE for SWS and PR.

| Parameter | Band | Default Bias | Wideband Bias | Default RMSE | Wideband RMSE |
|---|---|---|---|---|---|
| **SWS [m/s]** | **Low** | -0.281 | -0.019 | 0.312 | **0.038** |
| **SWS [m/s]** | **Mid** | -0.014 | 0.017 | 0.053 | **0.018** |
| **SWS [m/s]** | **High** | 0.060 | 0.062 | **0.061** | 0.064 |
| **PR [m/s]** | **Low** | -0.038 | 0.053 | **0.036** | 0.068 |
| **PR [m/s]** | **Mid** | 0.074 | 0.012 | 0.068 | **0.047** |
| **PR [m/s]** | **High** | 0.022 | -0.017 | 0.051 | **0.024** |

**Supplemental Table S3:** Shear wave speed (SWS), penetration rate (PR), and loss angle (ϕ) in healthy volunteers and patients within the low- (5-20 Hz), mid- (25-40 Hz), and high-

frequency (50-80 Hz) bands reported as median [IQR]. Group differences were assessed using the Wilcoxon rank-sum test.

| | | Liver | | | Spleen | | |
|---|---|---|---|---|---|---|---|
| **Parameter** | **Band** | **Healthy** | **Patient** | **p-value** | **Healthy** | **Patient** | **p-value** |
| **SWS [m/s]** | **Low** | 1.05 [1.03, 1.06] | 1.13 [1.08, 1.16] | **0.008** | 1.36 [1.29, 1.38] | 1.40 [1.35, 1.45] | 0.190 |
| **SWS [m/s]** | **Mid** | 1.22 [1.19, 1.25] | 1.33 [1.23, 1.46] | 0.063 | 1.90 [1.83, 1.94] | 1.96 [1.89, 2.00] | 0.258 |
| **SWS [m/s]** | **High** | 1.61 [1.56, 1.67] | 1.54 [1.49, 1.66] | 0.258 | 2.11 [2.05, 2.18] | 2.13 [2.05, 2.17] | 1.000 |
| **PR [m/s]** | **Low** | 0.59 [0.53, 0.64] | 0.37 [0.33, 0.50] | **0.003** | 0.43 [0.30, 0.53] | 0.38 [0.34, 0.42] | 0.297 |
| **PR [m/s]** | **Mid** | 1.17 [1.12, 1.24] | 0.94 [0.86, 1.18] | 0.113 | 1.32 [0.96, 1.61] | 1.28 [1.12, 1.38] | 0.863 |
| **PR [m/s]** | **High** | 1.45 [1.35, 1.50] | 1.43 [1.27, 1.57] | 0.931 | 2.14 [1.70, 2.42] | 2.49 [2.35, 2.83] | 0.113 |
| **ϕ [rad]** | **Low** | 0.60 [0.59, 0.69] | 0.99 [0.80, 1.04] | **<0.001** | 0.97 [0.86, 1.24] | 1.11 [1.00, 1.21] | 0.258 |
| **ϕ [rad]** | **Mid** | 0.33 [0.32, 0.35] | 0.43 [0.36, 0.52] | **0.024** | 0.48 [0.40, 0.60] | 0.52 [0.46, 0.61] | 0.387 |
| **ϕ [rad]** | **High** | 0.34 [0.34, 0.36] | 0.34 [0.33, 0.37] | 0.546 | 0.31 [0.29, 0.38] | 0.27 [0.25, 0.29] | 0.113 |

**Supplemental Note S2: Phase-gradient consistency mapping and regional phase unwrapping**

Spatial derivatives used by k-MDEV are sensitive to residual $2\pi$ phase discontinuities. In abdominal MRE, strong motion and low signal outside the target organs can generate phase residues whose influence is distributed by global least-squares Laplacian unwrapping and may extend into the liver or spleen (Figure S4A). To reduce this coupling, a preliminary global Laplacian solution was used to derive a local phase-gradient consistency map, after which iterative Laplacian phase unwrapping (ILPU) [32] was applied independently within organ-specific regions.

**Phase-gradient consistency map.** Let $x = (i, j)$ denote an in-plane voxel and let $s$ index the phase offset, motion-encoding-gradient (MEG) component, and driver frequency. The measured wrapped phase is $\phi_s^w(x) \in (-\pi, \pi]$, and $\tilde{\phi}_s(x)$ denotes the preliminary phase obtained by single-step two-dimensional Laplacian unwrapping with Neumann boundary conditions. The wrapping operator was defined as

$$W(a) = \mathrm{atan2}(\sin(a, \cos a) \in (-\pi, \pi].$$

For a valid nearest-neighbor edge $e = (x, y)$, with $D_e\phi = \phi(y) - \phi(x)$, the circular discrepancy between the measured and preliminary wrapped phase gradients was calculated as

$$d_{e,s} = \left| W\left( W(D_e\phi_s^w) - W\left(D_e\tilde{\phi}_s\right)\right) \right| \in [0, \pi].$$

The voxel-wise consistency score was obtained by averaging over all valid nearest-neighbor edges incident to $x$,

$$q_s(x) = 1 - \frac{1}{\pi\, |\mathcal{N}_s(x)|} \sum_{y \in \mathcal{N}_s(x)} d_{(x,y),s} \in [0,1],$$

where $\mathcal{N}_s(x)$ contains only neighbors within the valid image support. The acquisition-averaged map was then calculated as

$$\overline{q}(x) = \frac{1}{|\mathcal{S}_x|} \sum_{s \in \mathcal{S}_x} q_s\,(x),$$

where $\mathcal{S}_x$ comprises the available phase offsets, MEG components, and driver frequencies at $x$. Higher values indicate closer local agreement between the measured wrapped gradients and those of the preliminary Laplacian solution. For organ $o$, the regional processing mask was defined by

$$M_o(x) = M_{\mathrm{ROI},o}(x)\, \mathbf{1}\{\overline{q}(x) \geq \tau_Q\},$$

where $M_{\mathrm{ROI},o}$ is the manually delineated liver or spleen mask and $\tau_Q$ is the consistency threshold.

Because both gradients in $d_{e,s}$ are rewrapped, $q_s = 1$ for any phase estimate that exactly preserves the measured wrapped gradients, including an estimate with an incorrect integer branch. The map was therefore interpreted as a local rewrapping-consistency measure for the preliminary Laplacian solution, rather than as independent evidence of correct integer-branch recovery.

**Regional phase unwrapping.** Each connected component of $M_0$ was processed independently with ILPU [32], thereby removing regularization links between the target organ and disconnected or unreliable abdominal regions. Before the component-wise solutions were combined, the additive constants arising from the Neumann Poisson problem were aligned to a common reference and the same convention was maintained across the phase-offset series. The regional ILPU values were subsequently held fixed as Dirichlet boundary data in a final Laplacian solve, which recovered the remaining voxels within the abdominal region of interest and yielded a globally defined unwrapped phase image (Figure 1). This regional construction limits residue spreading across disconnected tissue supports; however, a component-wide integer offset remains unidentifiable from spatial gradients alone and must be kept consistent before temporal Fourier transformation.

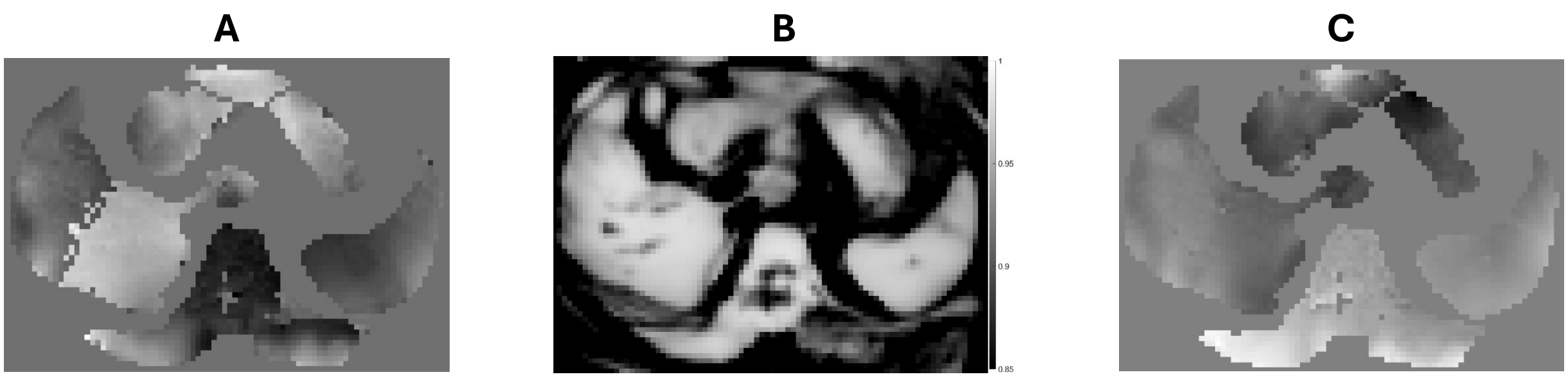

**Supplemental Figure S2:** Representative abdominal phase unwrapping. (A) Single-step global Laplacian unwrapping, showing a phase discontinuity extending into the liver. (B) Phase-gradient consistency map averaged over phase offsets, MEG components, and driver frequencies; higher values indicate closer local agreement between measured and preliminary wrapped gradients. (C) Phase after regional ILPU. The same axial slice is shown in all panels; the grayscale ranges of the two phase images were normalized independently for display.

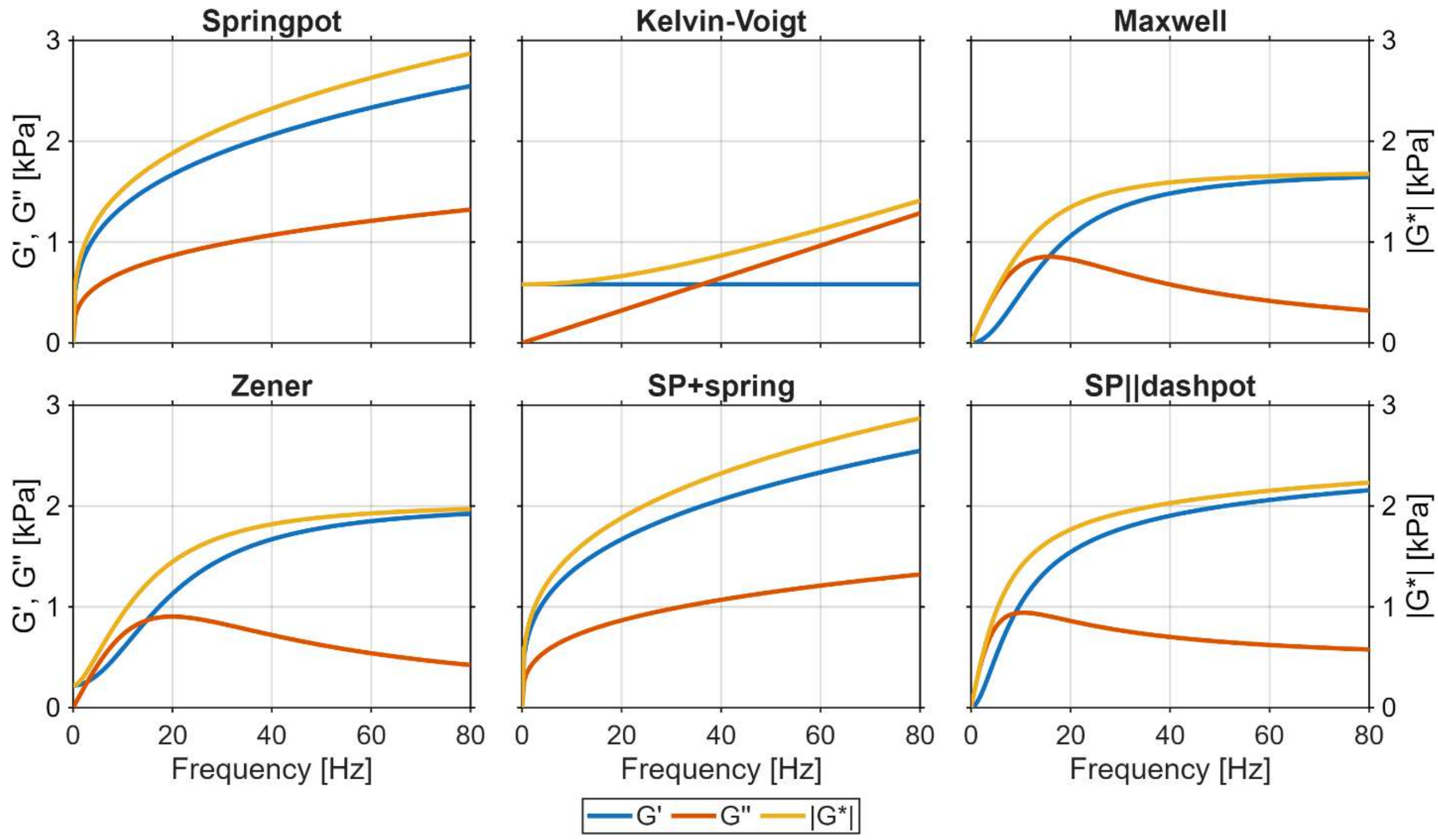

**Supplemental Figure S3:** Model-predicted dispersion of storage modulus (G’, blue), loss modulus (G’’, orange), and the magnitude of the complex shear modulus (|G*|, yellow) in liver tissue.

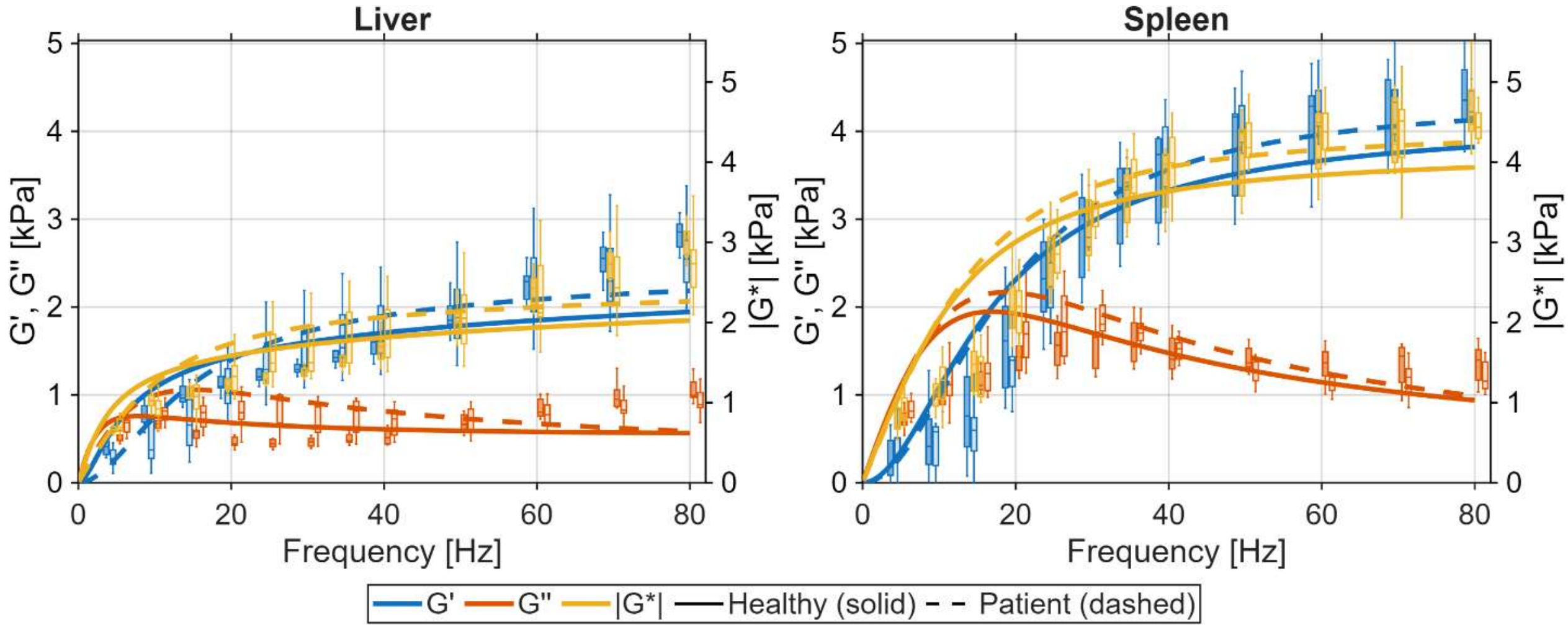

**Supplemental Figure S4:** Measured and SP-dashpot model-predicted dispersion of storage modulus (G’, blue), loss modulus (G’’, orange), and the magnitude of the complex shear modulus (|G*|, yellow) in liver and spleen tissue.